\documentclass[prl,twocolumn,superscriptaddress]{revtex4}
\usepackage{graphicx}
\usepackage{comment}
\usepackage{amsmath,amssymb,amsthm} 
\usepackage{verbatim}
\usepackage{float}

\newcommand{\be}{\begin{equation}}
\newcommand{\ee}{\end{equation}}

\begin{document}

\title{Aging of CKN: Modulus versus conductivity analysis}
\author{Jeppe C. Dyre}
\affiliation{DNRF Centre ``Glass and Time'', IMFUFA, Department of Sciences, Roskilde University, Postbox 260, DK-4000 Roskilde, Denmark}

\date{\today}

\begin{abstract}
It was recently reported that the electrical modulus peaks narrows upon annealing of the ionic system CKN [Paluch {\it et al.}, Phys. Rev. Lett. {\bf 110}, 015702 (2013)], which was interpreted as providing evidence of dynamic heterogeneity of this glass-forming liquid. An analysis of the same data in terms of the ac conductivity shows no shape changes, however. We discuss the relation between both findings and show further that the ac conductivity conforms to the prediction of the random barrier model (RBM) at all times during the annealing.
\end{abstract}

\maketitle

Ionic conduction in glasses and other disordered solids is a subject of growing interest due to applications in connection with solid-oxide fuel cells, electrochemical sensors, thin-film solid electrolytes in batteries and supercapacitors, electrochromic windows, oxygen-separation membranes, functional polymers, etc. At the same time ion conduction in disordered solids remains an area of basic research because a number of fundamental questions are still not settled \cite{ing87,ang90,mai95,dyr08}. 

In a recent Letter Paluch, Wojnarowska, and Hensel-Bielowka \cite{pal13} presented data for (physical) aging of the ionic glass ${\rm [Ca(NO_3)_2]_{0.4}[KNO_3]_{0.6}}$ (CKN), for which the glass transition temperature is 335 K. Liquid CKN is an ionic glass former with a significant decoupling of the ionic motion from the structural relaxation, implying much faster ionic motion than estimated from the liquid's viscosity via the Stokes-Einstein relation \cite{pim96}. In other words, the conductivity relaxation time of liquid CKN is much shorter than the Maxwell relaxation time. This is in contrast to the recently intensely studied room temperature ionic liquids, for which anions and cations are of roughly the same size and, consequently, little or no decoupling is observed \cite{san09,san11}.

The decoupling of ionic motion from the structural relaxation in CKN makes it simple to monitor aging by measuring the frequency-dependent conductivity. For most glass-forming liquids the alpha dielectric loss peak is closely linked to the structural relaxation \cite{jac12}, making it impossible to measure the entire alpha dielectric relaxation process under constant conditions in the glass phase because the sample changes its properties due to physical aging. Actually, aging has been studied by ac methods even for such glasses -- either by monitoring the dielectric loss at much higher frequencies than the alpha loss-peak frequency \cite{joh82,ale95,leh98,can05,lun05,hec10} or by monitoring the beta process \cite{joh82,dyr03} -- but studying aging of CKN by ac methods is conceptually much simpler. 

Recall that if $\omega$ is the angular frequency and $\sigma(\omega)$ the frequency-dependent complex conductivity, the complex electrical modulus $M(\omega)=M'(\omega)+iM''(\omega)$ is defined \cite{mac72} by 

\be\label{m}
M(\omega)
\equiv \varepsilon_0\frac{i\omega}{\sigma(\omega)}\,.
\ee
Based on the observed narrowing of the loss modulus $M''(\omega)$ for CKN (compare Fig. 1(b)) and other ionic glasses the authors of Ref. \onlinecite{pal13} concluded that ``... the changes in the conductivity relaxation process observed during isothermal aging ... provide strong experimental evidence of the heterogeneous nature of deeply supercooled liquids.'' 

Some time ago there was an engaged debate in the literature about which method of data representation -- modulus or conductivity -- yields most insight into the physics of ionic conductors \cite{alm83,dyr91,ell94,rol98,nga00,sid00}. Below we present an analysis of the CKN data of Ref. \onlinecite{pal13} from the conductivity viewpoint and argue that no conclusion can be drawn about the absence or presence of dynamic heterogeneities. We show further that the data conform to the prediction of the random barrier model (RBM), a simple effectively zero-parameter model of ionic motion in highly disordered structures.

The CKN sample of Ref. \onlinecite{pal13} was first annealed at 353 K for 10 minutes, i.e., much above the glass transition temperature (335 K), and then  quenched to 308 K at a rate of 10-15 K/min. Hereafter temperature was kept constant and $\sigma(\omega)$ was measured over the frequency range $0.01-10^6$ Hz every 15 minutes. Figure 1(a) shows a log-log plot of two sets of data for $M''(\omega)$, one set obtained two hours after the glass was produced (red) and one set obtained 22 hours after (blue). Figure 1(b) shows the same data scaled to make the two maxima coincide. Though the effect is not large, annealing clearly leads to a narrowing of the modulus peak \cite{rol98,how74}. If the peak is decomposed mathematically as a sum of Debye peaks, the corresponding relaxation time distribution narrows upon annealing.

\begin{figure}[H]
  \centering
  \includegraphics[width=80mm]{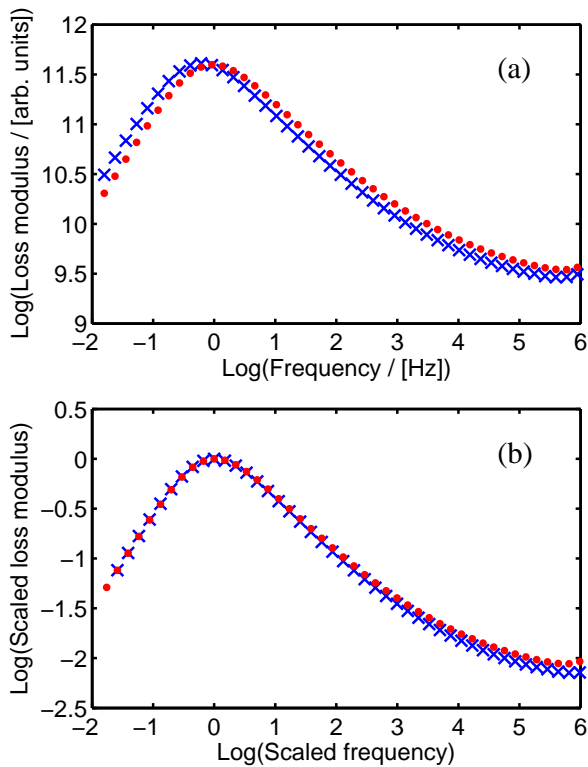}
  \caption{(a) Log-log plot (base 10) of the loss modulus of CKN annealed at temperature 308 K. The figure shows data measured two hours after cooling to 308 K at a rate of 10-15 K/min from above the glass transition at 335 K (red dots), as well as 22 hours after cooling to 308 K (blue crosses) (data from Ref. \onlinecite{pal13}).
(b) The same data plotted by scaling with the modulus maximum and the corresponding frequency. The modulus peak narrows upon annealing.}
  \label{fig1}
\end{figure}

Figure 2 shows the conductivity analysis of the same data. In Fig. 2(a) we plot the real part of the conductivity as a function of frequency in a log-log plot. Upon annealing the conductivity decreases at all frequencies. One possible explanation could be that the density increases upon annealing, making it more difficult for the ions to move; however, we do not wish to speculate here about what is the physical mechanism behind the conductivity decrease \cite{rol98}. Figure 2(b) shows the same data relative to the dc conductivity plotted as a function of frequency scaled empirically to obtain the best overlap between the two curves. No shape change is observed. Such behavior is often observed in physical aging experiments, where it is referred to as time aging-time superposition \cite{str78,oco99,hec10}. If interpreted in terms of a relaxation time distribution, time aging-time superposition implies that the distribution does not narrow upon annealing. How is one to understand this, given that the modulus relaxation time distribution does narrow?

\begin{figure}[H]
  \centering
  \includegraphics[width=80mm]{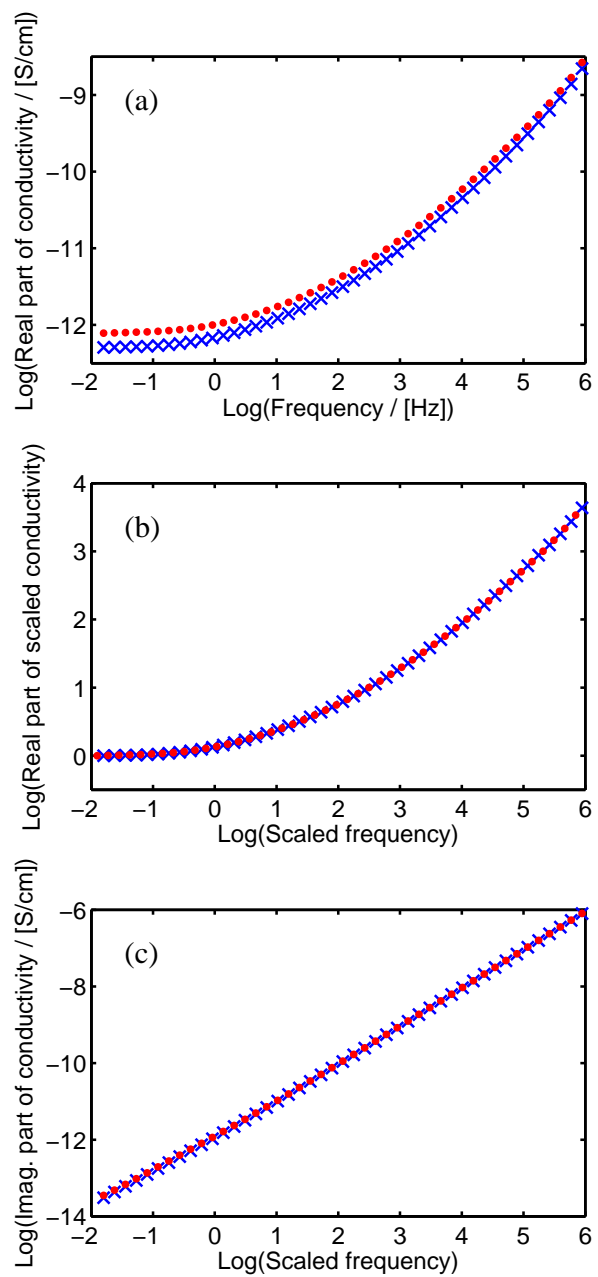}
  \caption{(a) An alternative representation of the data of Fig. 1 showing the real part of the ac conductivity $\sigma(\omega)$ of CKN annealed at 308 K two hours after the glass transition (red dots) and 22 hours after the glass transition (blue crosses). It is seen that annealing decreases the conductivity at all frequencies.
(b) The real part of the scaled ac conductivity $\sigma(\omega)/\sigma(0)$ two hours after the glass transition (red dots) and 22 hours after (blue crosses) plotted as a function of frequency scaled empirically to maximize the overlap. No shape change is observed.
(c) The imaginary part of the conductivity two hours after the glass transition (red dots) and 22 hours after (blue crosses). Virtually no changes are observed, and for both data sets the imaginary part is almost proportional to frequency. These facts show that the major contribution to the imaginary part derives from the electronic polarization (which is instantaneous compared to the ionic motion), an observation that is key to understanding why the modulus peak narrows during annealing (Fig. 1).}
  \label{fig2}
\end{figure}

A clue is provided by the imaginary part of the ac conductivity plotted in Fig. 2(c). Note that 1) there is virtually no change upon annealing; 2) the imaginary part of $\sigma(\omega)$ is almost proportional to frequency; 3) at most frequencies the imaginary part is much larger than the real part. These observations show that the imaginary part of the ac conductivity over most of the frequency range monitored is dominated by the charge displacements coming from the instantaneous electronic polarization, which is quantified by the high-frequency dielectric constant $\epsilon_\infty$. The electronic polarization is independent of the ion motion monitored by the real part of the conductivity. Because the conductivity appears in the denominator in Eq. (\ref{m}), however, ion motion and $\epsilon_\infty$ polarization {\it both} influence $M''(\omega)$. In fact, if $\epsilon_\infty$ is constant during annealing and the real part of conductivity decreases -- as observed -- the mathematics implies that the modulus peak {\it must} narrow \cite{dyr91,sid00}.

\begin{figure}[H]
  \centering
  \includegraphics[width=90mm]{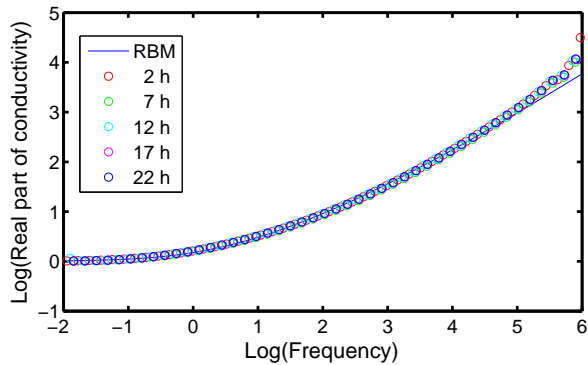}
  \caption{Real part of $\sigma(\omega)/\sigma(0)$ plotted as a function of empirically scaled frequency for CKN annealed at 308K. The figure shows data taken 2, 7, 12, 17, and 22 hours after the glass transition. No shape change is observed. The full curve is the prediction of the random barrier model (RBM), Eq. (\ref{rbm}).
} \label{fig3}
\end{figure}

Figure 3 shows data for the real part of the scaled ac conductivity $\sigma(\omega)/\sigma(0)$ taken at five different times during the annealing at 308 K; the frequencies have been scaled empirically for best overlap in order to investigate whether or not the shape changes. The full curve is the prediction of the random barrier model (RBM) \cite{dyr88,sch00}, a simple model for ac conduction in disordered solids based on the following assumptions: 1) All ion-ion interactions including self-exclusion are ignored; 2) the ion sites have the same energy; 3) the ions move on a simple cubic lattice where only nearest-neighbor jumps are allowed; 4) the jump rates are determined by energy barriers that vary randomly and spatically uncorrelated. In the extreme disorder limit, i.e., when the width of the energy barrier distribution is much larger than $k_BT$, the scaled ac conductivity is independent of the barrier distribution \cite{sch00} -- in this limit both dc and ac conduction are dominated by percolation \cite{sch08}. In the extreme disorder limit the RBM is to a very good approximation described by the following equation for $\tilde\sigma\equiv\sigma(\omega)/\sigma(0)$ 

\be\label{rbm}
\ln\tilde\sigma
\,=\,\left(\frac{i\tilde\omega}{\tilde\sigma}\right)^{2/3}\,.
\ee
In the derivation of this equation \cite{sch08} the number $2/3$ appears because it is half the exponent $4/3$ of the Alexander-Orbach conjecture \cite{ale82} for the spectral dimension of the percolation cluster, a conjecture that is known to be almost correct in any number of dimensions \cite{nak94,hug96}. Only at low frequencies where the conductivity is almost constant does Eq. (2) become inaccurate in describing the difference $\tilde\sigma(\tilde\omega)-1$ of the RBM; in this frequency range a more accurate approximate analytical expression for the ac conductivity is available \cite{sch08}.

In summary, we have shown that a conductivity-based analysis of data for the ionic glass former CKN annealed below the glass transition temperature reveals no shape changes, i.e., the conductivity obeys time aging-time superposition. As time progresses the entire real part of the conductivity is displaced to lower values. This may reflect the sample compactifying slightly, but may also derive from a change in the effective number of mobile ions \cite{dyr08,rol98}. In any case, from the conductivity viewpoint the data do not support the interpretation of Ref. \onlinecite{pal13}, according to which the measurements confirm ``the existence of slow and fast subensembles in the glassy state [which] should also result in distinct aging rates for these different regions in the system.'' In fact, the RBM fits data well at all times during the annealing, which shows that there is no need to invoke changes in the degree of static or dynamic heterogenities. -- 
We do not wish suggest that viscous liquids and glasses are dynamically homogeneous \cite{boh96,kob97,edi00,vol02,ber11b,dynhet}. It seems difficult to imagine a realistic model of a disordered system with energy barriers that are large compared to $k_BT$ without some sort of dynamic heterogeneity; for instance the RBM is a model for which spatial inhomogeneities are crucially important. The fundamental question is not whether dynamic heterogeneities exist, but whether these {\it cause} the observed physics or are an {\it effect} of the disorder, which is particularly pronounced for ultraviscous liquids and glasses.

\acknowledgments 

The author is indebted to Marian Paluch for making available the CKN data and to Tina Hecksher for technical assistance. 
The center for viscous liquid dynamics ``Glass and Time'' is sponsored by the Danish National Research Foundation via grant DNRF61.


\begin{thebibliography}{38}
\expandafter\ifx\csname natexlab\endcsname\relax\def\natexlab#1{#1}\fi
\expandafter\ifx\csname bibnamefont\endcsname\relax
  \def\bibnamefont#1{#1}\fi
\expandafter\ifx\csname bibfnamefont\endcsname\relax
  \def\bibfnamefont#1{#1}\fi
\expandafter\ifx\csname citenamefont\endcsname\relax
  \def\citenamefont#1{#1}\fi
\expandafter\ifx\csname url\endcsname\relax
  \def\url#1{\texttt{#1}}\fi
\expandafter\ifx\csname urlprefix\endcsname\relax\def\urlprefix{URL }\fi
\providecommand{\bibinfo}[2]{#2}
\providecommand{\eprint}[2][]{\url{#2}}

\bibitem[{\citenamefont{Ingram}(1987)}]{ing87}
\bibinfo{author}{\bibfnamefont{M.~D.} \bibnamefont{Ingram}},
  \bibinfo{journal}{Phys. Chem. Glasses} \textbf{\bibinfo{volume}{28}},
  \bibinfo{pages}{215} (\bibinfo{year}{1987}).

\bibitem[{\citenamefont{Angell}(1990)}]{ang90}
\bibinfo{author}{\bibfnamefont{C.~A.} \bibnamefont{Angell}},
  \bibinfo{journal}{Chem. Rev.} \textbf{\bibinfo{volume}{90}},
  \bibinfo{pages}{523} (\bibinfo{year}{1990}).

\bibitem[{\citenamefont{Maier}(1995)}]{mai95}
\bibinfo{author}{\bibfnamefont{J.}~\bibnamefont{Maier}},
  \bibinfo{journal}{Prog. Solid State Chem.} \textbf{\bibinfo{volume}{23}},
  \bibinfo{pages}{171} (\bibinfo{year}{1995}).

\bibitem[{\citenamefont{Dyre et~al.}(2008)\citenamefont{Dyre, Maass, Roling,
  and Sidebottom}}]{dyr08}
\bibinfo{author}{\bibfnamefont{J.~C.} \bibnamefont{Dyre}},
  \bibinfo{author}{\bibfnamefont{P.}~\bibnamefont{Maass}},
  \bibinfo{author}{\bibfnamefont{B.}~\bibnamefont{Roling}}, \bibnamefont{and}
  \bibinfo{author}{\bibfnamefont{D.~L.} \bibnamefont{Sidebottom}},
  \bibinfo{journal}{Rep. Prog. Phys.} \textbf{\bibinfo{volume}{72}},
  \bibinfo{pages}{046501} (\bibinfo{year}{2008}).

\bibitem[{\citenamefont{Paluch et~al.}(2013)\citenamefont{Paluch, Wojnarowska,
  and Hensel-Bielowka}}]{pal13}
\bibinfo{author}{\bibfnamefont{M.}~\bibnamefont{Paluch}},
  \bibinfo{author}{\bibfnamefont{Z.}~\bibnamefont{Wojnarowska}},
  \bibnamefont{and}
  \bibinfo{author}{\bibfnamefont{S.}~\bibnamefont{Hensel-Bielowka}},
  \bibinfo{journal}{Phys. Rev. Lett.} \textbf{\bibinfo{volume}{110}},
  \bibinfo{pages}{015702} (\bibinfo{year}{2013}).

\bibitem[{\citenamefont{Pimenov et~al.}(1996)\citenamefont{Pimenov,
  Lunkenheimer, Rall, Kohlhaas, Loidl, and B\"ohmer}}]{pim96}
\bibinfo{author}{\bibfnamefont{A.}~\bibnamefont{Pimenov}},
  \bibinfo{author}{\bibfnamefont{P.}~\bibnamefont{Lunkenheimer}},
  \bibinfo{author}{\bibfnamefont{H.}~\bibnamefont{Rall}},
  \bibinfo{author}{\bibfnamefont{R.}~\bibnamefont{Kohlhaas}},
  \bibinfo{author}{\bibfnamefont{A.}~\bibnamefont{Loidl}}, \bibnamefont{and}
  \bibinfo{author}{\bibfnamefont{R.}~\bibnamefont{B\"ohmer}},
  \bibinfo{journal}{Phys. Rev. E} \textbf{\bibinfo{volume}{54}},
  \bibinfo{pages}{676} (\bibinfo{year}{1996}).

\bibitem[{\citenamefont{Sangoro et~al.}(2009)\citenamefont{Sangoro, Iacob,
  Serghei, Friedrich, and Kremer}}]{san09}
\bibinfo{author}{\bibfnamefont{J.~R.} \bibnamefont{Sangoro}},
  \bibinfo{author}{\bibfnamefont{C.}~\bibnamefont{Iacob}},
  \bibinfo{author}{\bibfnamefont{A.}~\bibnamefont{Serghei}},
  \bibinfo{author}{\bibfnamefont{C.}~\bibnamefont{Friedrich}},
  \bibnamefont{and} \bibinfo{author}{\bibfnamefont{F.}~\bibnamefont{Kremer}},
  \bibinfo{journal}{Phys. Chem. Chem. Phys.} \textbf{\bibinfo{volume}{11}},
  \bibinfo{pages}{913} (\bibinfo{year}{2009}).

\bibitem[{\citenamefont{Sangoro et~al.}(2011)\citenamefont{Sangoro, Iacob,
  Naumov, Valiullin, Rexhausen, Hunger, Buchner, Strehmel, K{\"a}rger, and
  Kremer}}]{san11}
\bibinfo{author}{\bibfnamefont{J.~R.} \bibnamefont{Sangoro}},
  \bibinfo{author}{\bibfnamefont{C.}~\bibnamefont{Iacob}},
  \bibinfo{author}{\bibfnamefont{S.}~\bibnamefont{Naumov}},
  \bibinfo{author}{\bibfnamefont{R.}~\bibnamefont{Valiullin}},
  \bibinfo{author}{\bibfnamefont{H.}~\bibnamefont{Rexhausen}},
  \bibinfo{author}{\bibfnamefont{J.}~\bibnamefont{Hunger}},
  \bibinfo{author}{\bibfnamefont{R.}~\bibnamefont{Buchner}},
  \bibinfo{author}{\bibfnamefont{V.}~\bibnamefont{Strehmel}},
  \bibinfo{author}{\bibfnamefont{J.}~\bibnamefont{K{\"a}rger}},
  \bibnamefont{and} \bibinfo{author}{\bibfnamefont{F.}~\bibnamefont{Kremer}},
  \bibinfo{journal}{Soft Matter} \textbf{\bibinfo{volume}{7}},
  \bibinfo{pages}{1678} (\bibinfo{year}{2011}).

\bibitem[{\citenamefont{Jakobsen et~al.}(2012)\citenamefont{Jakobsen, Hecksher,
  Christensen, Olsen, Dyre, and Niss}}]{jac12}
\bibinfo{author}{\bibfnamefont{B.}~\bibnamefont{Jakobsen}},
  \bibinfo{author}{\bibfnamefont{T.}~\bibnamefont{Hecksher}},
  \bibinfo{author}{\bibfnamefont{T.}~\bibnamefont{Christensen}},
  \bibinfo{author}{\bibfnamefont{N.~B.} \bibnamefont{Olsen}},
  \bibinfo{author}{\bibfnamefont{J.~C.} \bibnamefont{Dyre}}, \bibnamefont{and}
  \bibinfo{author}{\bibfnamefont{K.}~\bibnamefont{Niss}}, \bibinfo{journal}{J.
  Chem. Phys.} \textbf{\bibinfo{volume}{136}}, \bibinfo{pages}{081102}
  (\bibinfo{year}{2012}).

\bibitem[{\citenamefont{Johari}(1982)}]{joh82}
\bibinfo{author}{\bibfnamefont{G.~P.} \bibnamefont{Johari}},
  \bibinfo{journal}{J. Chem. Phys.} \textbf{\bibinfo{volume}{77}},
  \bibinfo{pages}{4619} (\bibinfo{year}{1982}).

\bibitem[{\citenamefont{Alegria et~al.}(1995)\citenamefont{Alegria,
  Guerrica-Echevarria, Goitiandia, Telleria, and Colmenero}}]{ale95}
\bibinfo{author}{\bibfnamefont{A.}~\bibnamefont{Alegria}},
  \bibinfo{author}{\bibfnamefont{E.}~\bibnamefont{Guerrica-Echevarria}},
  \bibinfo{author}{\bibfnamefont{L.}~\bibnamefont{Goitiandia}},
  \bibinfo{author}{\bibfnamefont{I.}~\bibnamefont{Telleria}}, \bibnamefont{and}
  \bibinfo{author}{\bibfnamefont{J.}~\bibnamefont{Colmenero}},
  \bibinfo{journal}{Macromolecules} \textbf{\bibinfo{volume}{28}},
  \bibinfo{pages}{1516} (\bibinfo{year}{1995}).

\bibitem[{\citenamefont{Leheny and Nagel}(1998)}]{leh98}
\bibinfo{author}{\bibfnamefont{R.~L.} \bibnamefont{Leheny}} \bibnamefont{and}
  \bibinfo{author}{\bibfnamefont{S.~R.} \bibnamefont{Nagel}},
  \bibinfo{journal}{Phys. Rev. B} \textbf{\bibinfo{volume}{57}},
  \bibinfo{pages}{5154} (\bibinfo{year}{1998}).

\bibitem[{\citenamefont{Cangialosi et~al.}(2005)\citenamefont{Cangialosi,
  W{\"u}bbenhorst, Groenewold, Mendes, and Picken}}]{can05}
\bibinfo{author}{\bibfnamefont{D.}~\bibnamefont{Cangialosi}},
  \bibinfo{author}{\bibfnamefont{M.}~\bibnamefont{W{\"u}bbenhorst}},
  \bibinfo{author}{\bibfnamefont{J.}~\bibnamefont{Groenewold}},
  \bibinfo{author}{\bibfnamefont{E.}~\bibnamefont{Mendes}}, \bibnamefont{and}
  \bibinfo{author}{\bibfnamefont{S.~J.} \bibnamefont{Picken}},
  \bibinfo{journal}{J. Non-Cryst. Solids} \textbf{\bibinfo{volume}{351}},
  \bibinfo{pages}{2605} (\bibinfo{year}{2005}).

\bibitem[{\citenamefont{Lunkenheimer et~al.}(2005)\citenamefont{Lunkenheimer,
  Wehn, Schneider, and Loidl}}]{lun05}
\bibinfo{author}{\bibfnamefont{P.}~\bibnamefont{Lunkenheimer}},
  \bibinfo{author}{\bibfnamefont{R.}~\bibnamefont{Wehn}},
  \bibinfo{author}{\bibfnamefont{U.}~\bibnamefont{Schneider}},
  \bibnamefont{and} \bibinfo{author}{\bibfnamefont{A.}~\bibnamefont{Loidl}},
  \bibinfo{journal}{Phys. Rev. Lett.} \textbf{\bibinfo{volume}{95}},
  \bibinfo{pages}{055702} (\bibinfo{year}{2005}).

\bibitem[{\citenamefont{Hecksher et~al.}(2010)\citenamefont{Hecksher, Olsen,
  Niss, and Dyre}}]{hec10}
\bibinfo{author}{\bibfnamefont{T.}~\bibnamefont{Hecksher}},
  \bibinfo{author}{\bibfnamefont{N.~B.} \bibnamefont{Olsen}},
  \bibinfo{author}{\bibfnamefont{K.}~\bibnamefont{Niss}}, \bibnamefont{and}
  \bibinfo{author}{\bibfnamefont{J.~C.} \bibnamefont{Dyre}},
  \bibinfo{journal}{J. Chem. Phys.} \textbf{\bibinfo{volume}{133}},
  \bibinfo{pages}{174514} (\bibinfo{year}{2010}).

\bibitem[{\citenamefont{Dyre and Olsen}(2003)}]{dyr03}
\bibinfo{author}{\bibfnamefont{J.~C.} \bibnamefont{Dyre}} \bibnamefont{and}
  \bibinfo{author}{\bibfnamefont{N.~B.} \bibnamefont{Olsen}},
  \bibinfo{journal}{Phys. Rev. Lett.} \textbf{\bibinfo{volume}{91}},
  \bibinfo{pages}{155703} (\bibinfo{year}{2003}).

\bibitem[{\citenamefont{Macedo et~al.}(1972)\citenamefont{Macedo, Moynihan, and
  Bose}}]{mac72}
\bibinfo{author}{\bibfnamefont{P.~B.} \bibnamefont{Macedo}},
  \bibinfo{author}{\bibfnamefont{C.~T.} \bibnamefont{Moynihan}},
  \bibnamefont{and} \bibinfo{author}{\bibfnamefont{R.}~\bibnamefont{Bose}},
  \bibinfo{journal}{Phys. Chem. Glasses} \textbf{\bibinfo{volume}{13}},
  \bibinfo{pages}{171} (\bibinfo{year}{1972}).

\bibitem[{\citenamefont{Almond and West}(1983)}]{alm83}
\bibinfo{author}{\bibfnamefont{D.~P.} \bibnamefont{Almond}} \bibnamefont{and}
  \bibinfo{author}{\bibfnamefont{A.~R.} \bibnamefont{West}},
  \bibinfo{journal}{Solid State lon.} \textbf{\bibinfo{volume}{11}},
  \bibinfo{pages}{57} (\bibinfo{year}{1983}).

\bibitem[{\citenamefont{Dyre}(1991)}]{dyr91}
\bibinfo{author}{\bibfnamefont{J.~C.} \bibnamefont{Dyre}}, \bibinfo{journal}{J.
  Non-Cryst. Solids} \textbf{\bibinfo{volume}{135}}, \bibinfo{pages}{219}
  (\bibinfo{year}{1991}).

\bibitem[{\citenamefont{Elliott}(1994)}]{ell94}
\bibinfo{author}{\bibfnamefont{S.~R.} \bibnamefont{Elliott}},
  \bibinfo{journal}{J. Non-Cryst. Solids} \textbf{\bibinfo{volume}{170}},
  \bibinfo{pages}{97} (\bibinfo{year}{1994}).

\bibitem[{\citenamefont{Roling}(1998)}]{rol98}
\bibinfo{author}{\bibfnamefont{B.}~\bibnamefont{Roling}},
  \bibinfo{journal}{Solid State Ion.} \textbf{\bibinfo{volume}{105}},
  \bibinfo{pages}{185} (\bibinfo{year}{1998}).

\bibitem[{\citenamefont{Ngai and Rendell}(2000)}]{nga00}
\bibinfo{author}{\bibfnamefont{K.~L.} \bibnamefont{Ngai}} \bibnamefont{and}
  \bibinfo{author}{\bibfnamefont{R.~W.} \bibnamefont{Rendell}},
  \bibinfo{journal}{Phys. Rev. B} \textbf{\bibinfo{volume}{61}},
  \bibinfo{pages}{9393} (\bibinfo{year}{2000}).

\bibitem[{\citenamefont{Sidebottom et~al.}(2000)\citenamefont{Sidebottom,
  Roling, and Funke}}]{sid00}
\bibinfo{author}{\bibfnamefont{D.~L.} \bibnamefont{Sidebottom}},
  \bibinfo{author}{\bibfnamefont{B.}~\bibnamefont{Roling}}, \bibnamefont{and}
  \bibinfo{author}{\bibfnamefont{K.}~\bibnamefont{Funke}},
  \bibinfo{journal}{Phys. Rev. B} \textbf{\bibinfo{volume}{63}},
  \bibinfo{pages}{024301} (\bibinfo{year}{2000}).

\bibitem[{\citenamefont{Howell et~al.}(1974)\citenamefont{Howell, Bose, Macedo,
  and Moynihan}}]{how74}
\bibinfo{author}{\bibfnamefont{F.~S.} \bibnamefont{Howell}},
  \bibinfo{author}{\bibfnamefont{R.~A.} \bibnamefont{Bose}},
  \bibinfo{author}{\bibfnamefont{P.~B.} \bibnamefont{Macedo}},
  \bibnamefont{and} \bibinfo{author}{\bibfnamefont{C.~T.}
  \bibnamefont{Moynihan}}, \bibinfo{journal}{J. Phys. Chem.}
  \textbf{\bibinfo{volume}{78}}, \bibinfo{pages}{639} (\bibinfo{year}{1974}).

\bibitem[{\citenamefont{Struik}(1978)}]{str78}
\bibinfo{author}{\bibfnamefont{L.~C.~E.} \bibnamefont{Struik}},
  \emph{\bibinfo{title}{Physical Aging in Amorphous Polymers and Other
  Materials}} (\bibinfo{publisher}{Elsevier, Amsterdam}, \bibinfo{year}{1978}).

\bibitem[{\citenamefont{O'Connell and B.McKenna}(1999)}]{oco99}
\bibinfo{author}{\bibfnamefont{P.~A.} \bibnamefont{O'Connell}}
  \bibnamefont{and}
  \bibinfo{author}{\bibfnamefont{G.}~\bibnamefont{B.McKenna}},
  \bibinfo{journal}{J. Chem. Phys.} \textbf{\bibinfo{volume}{110}},
  \bibinfo{pages}{11054} (\bibinfo{year}{1999}).

\bibitem[{\citenamefont{Dyre}(1988)}]{dyr88}
\bibinfo{author}{\bibfnamefont{J.~C.} \bibnamefont{Dyre}}, \bibinfo{journal}{J.
  Appl. Phys.} \textbf{\bibinfo{volume}{64}}, \bibinfo{pages}{2456}
  (\bibinfo{year}{1988}).

\bibitem[{\citenamefont{Schr{\o}der and Dyre}(2000)}]{sch00}
\bibinfo{author}{\bibfnamefont{T.~B.} \bibnamefont{Schr{\o}der}}
  \bibnamefont{and} \bibinfo{author}{\bibfnamefont{J.~C.} \bibnamefont{Dyre}},
  \bibinfo{journal}{Rev. Mod. Phys.} \textbf{\bibinfo{volume}{72}},
  \bibinfo{pages}{873} (\bibinfo{year}{2000}).

\bibitem[{\citenamefont{Schr{\o}der and Dyre}(2008)}]{sch08}
\bibinfo{author}{\bibfnamefont{T.~B.} \bibnamefont{Schr{\o}der}}
  \bibnamefont{and} \bibinfo{author}{\bibfnamefont{J.~C.} \bibnamefont{Dyre}},
  \bibinfo{journal}{Phys. Rev. Lett.} \textbf{\bibinfo{volume}{101}},
  \bibinfo{pages}{025901} (\bibinfo{year}{2008}).

\bibitem[{\citenamefont{Alexander and Orbach}(1982)}]{ale82}
\bibinfo{author}{\bibfnamefont{S.}~\bibnamefont{Alexander}} \bibnamefont{and}
  \bibinfo{author}{\bibfnamefont{R.}~\bibnamefont{Orbach}},
  \bibinfo{journal}{J. Phys. (Paris) Lett.} \textbf{\bibinfo{volume}{43}},
  \bibinfo{pages}{625} (\bibinfo{year}{1982}).

\bibitem[{\citenamefont{Nakayama et~al.}(1994)\citenamefont{Nakayama, Yakubo,
  and Orbach}}]{nak94}
\bibinfo{author}{\bibfnamefont{T.}~\bibnamefont{Nakayama}},
  \bibinfo{author}{\bibfnamefont{K.}~\bibnamefont{Yakubo}}, \bibnamefont{and}
  \bibinfo{author}{\bibfnamefont{R.~L.} \bibnamefont{Orbach}},
  \bibinfo{journal}{Rev. Mod. Phys.} \textbf{\bibinfo{volume}{66}},
  \bibinfo{pages}{381} (\bibinfo{year}{1994}).

\bibitem[{\citenamefont{Hughes}(1996)}]{hug96}
\bibinfo{author}{\bibfnamefont{B.~D.} \bibnamefont{Hughes}},
  \emph{\bibinfo{title}{Random Walks and Random Environments}}
  (\bibinfo{publisher}{Clarendon, Oxford}, \bibinfo{year}{1996}).

\bibitem[{\citenamefont{B{\"o}hmer et~al.}(1996)\citenamefont{B{\"o}hmer,
  Hinze, Diezemann, Geil, and Sillescu}}]{boh96}
\bibinfo{author}{\bibfnamefont{R.}~\bibnamefont{B{\"o}hmer}},
  \bibinfo{author}{\bibfnamefont{G.}~\bibnamefont{Hinze}},
  \bibinfo{author}{\bibfnamefont{G.}~\bibnamefont{Diezemann}},
  \bibinfo{author}{\bibfnamefont{B.}~\bibnamefont{Geil}}, \bibnamefont{and}
  \bibinfo{author}{\bibfnamefont{H.}~\bibnamefont{Sillescu}},
  \bibinfo{journal}{EPL} \textbf{\bibinfo{volume}{36}}, \bibinfo{pages}{55}
  (\bibinfo{year}{1996}).

\bibitem[{\citenamefont{Kob et~al.}(1997)\citenamefont{Kob, Donati, Plimpton,
  Poole, and Glotzer}}]{kob97}
\bibinfo{author}{\bibfnamefont{W.}~\bibnamefont{Kob}},
  \bibinfo{author}{\bibfnamefont{C.}~\bibnamefont{Donati}},
  \bibinfo{author}{\bibfnamefont{S.~J.} \bibnamefont{Plimpton}},
  \bibinfo{author}{\bibfnamefont{P.~H.} \bibnamefont{Poole}}, \bibnamefont{and}
  \bibinfo{author}{\bibfnamefont{S.~C.} \bibnamefont{Glotzer}},
  \bibinfo{journal}{Phys. Rev. Lett.} \textbf{\bibinfo{volume}{79}},
  \bibinfo{pages}{2827} (\bibinfo{year}{1997}).

\bibitem[{\citenamefont{Ediger}(2000)}]{edi00}
\bibinfo{author}{\bibfnamefont{M.~D.} \bibnamefont{Ediger}},
  \bibinfo{journal}{Ann. Rev. Phys. Chem.} \textbf{\bibinfo{volume}{51}},
  \bibinfo{pages}{99} (\bibinfo{year}{2000}).

\bibitem[{\citenamefont{Vollmayr-Lee et~al.}(2002)\citenamefont{Vollmayr-Lee,
  Kob, Binder, and Zippelius}}]{vol02}
\bibinfo{author}{\bibfnamefont{K.}~\bibnamefont{Vollmayr-Lee}},
  \bibinfo{author}{\bibfnamefont{W.}~\bibnamefont{Kob}},
  \bibinfo{author}{\bibfnamefont{K.}~\bibnamefont{Binder}}, \bibnamefont{and}
  \bibinfo{author}{\bibfnamefont{A.}~\bibnamefont{Zippelius}},
  \bibinfo{journal}{J. Chem. Phys.} \textbf{\bibinfo{volume}{116}},
  \bibinfo{pages}{5158} (\bibinfo{year}{2002}).

\bibitem[{\citenamefont{Berthier and Biroli}(2011)}]{ber11b}
\bibinfo{author}{\bibfnamefont{L.}~\bibnamefont{Berthier}} \bibnamefont{and}
  \bibinfo{author}{\bibfnamefont{G.}~\bibnamefont{Biroli}},
  \bibinfo{journal}{Rev. Mod. Phys.} \textbf{\bibinfo{volume}{83}},
  \bibinfo{pages}{587} (\bibinfo{year}{2011}).

\bibitem[{\citenamefont{Berthier et~al.}(2011)\citenamefont{Berthier, Biroli,
  Bouchaud, Cipelletti, and van Saarloos}}]{dynhet}
\bibinfo{editor}{\bibfnamefont{L.}~\bibnamefont{Berthier}},
  \bibinfo{editor}{\bibfnamefont{G.}~\bibnamefont{Biroli}},
  \bibinfo{editor}{\bibfnamefont{J.-P.} \bibnamefont{Bouchaud}},
  \bibinfo{editor}{\bibfnamefont{L.}~\bibnamefont{Cipelletti}},
  \bibnamefont{and} \bibinfo{editor}{\bibfnamefont{W.}~\bibnamefont{van
  Saarloos}}, eds., \emph{\bibinfo{title}{Dynamical Heterogeneities in Glasses,
  Colloids, and Granular Media.}} (\bibinfo{publisher}{Oxford Univ. Press},
  \bibinfo{year}{2011}).

\end{thebibliography}



\end{document}